# Highly Boron-Doped Graphite and Diamond Synthesized From Adamantane and Ortho-Carborane under High Pressure

*Rustem H. Bagramov\*, Vladimir P. Filonenko, Igor P. Zibrov, Elena A. Skryleva, Alexander V. Nikolaev, Dmitrii G. Pasternak, Igor I. Vlasov*

This work demonstrates the effectiveness of the high-pressure method for the production of graphite and diamond with a high degree of boron doping using adamantane-carborane mixture as a precursor. At 8 GPa and 1700 °C, graphite is obtained from adamantane $C_{10}H_{16}$, whereas microcrystals of boron-doped diamond (2÷2.5 at.% of boron) are synthesized from a mixture of adamantane and ortho-carborane $C_2B_{10}H_{12}$ (atomic ratio B:C = 5:95). This result shows convincingly the catalytical activity of boron in the synthesis of diamond under high pressure. At pressures lower than 7 GPa, only graphite is synthesized from the adamantane and carborane mixture. Graphitization starts at quite low temperatures (below 1400 °C) and an increase in temperature simultaneously increases boron content and the quality of the graphite crystal lattice. Thorough study of the material structure allows us to assume that the substitutional boron atoms are distributed periodically and equidistantly from each other in the graphite layers at high boron concentrations (>1 at.%). The theoretical arguments and model ab initio calculations confirm this assumption and explain the experimentally observed boron concentrations.

**Introduction**

New carbon materials doped with boron have a unique set of physical properties, which opens up great prospects for them in high-tech applications.[1] For example, boron-doped graphene and graphite are promising for use in Li-ion batteries,[2] electrochemistry (various electrodes and sensors) and catalysts.[3-5] Boron-doping has been shown to improve the hydrogen storage capacity of graphite.[6, 7]

At a doping level of $10^{20} \div 10^{21}$ cm$^{-3}$, diamond has a metallic type of conductivity at elevated temperatures,[8-10] and above $10^{21}$ cm$^{-3}$ (1÷3 at.%) it is a superconductor at low temperatures.[11, 12] These characteristics, together with other physical and chemical properties, make up a good set that provides a wide range of new applications in electronic and electromechanical devices,[13] in electrodes,[14, 15] drilling and cutting tools,[16] and in other fields.[17]

There are known problems associated with the synthesis of highly boron-doped diamond. It is impossible to obtain it by the methods of ion implantation of boron, since this leads to graphitization of the diamond surface. It cannot be obtained by co-processing a pure diamond and a boron source under high pressure and temperature (HPHT) conditions. Boron can enter the diamond lattice during its formation, for example, when using the traditional method of high static pressures using catalyst metals. However, this method fails to achieve a high degree of boron doping both in large single crystals,[18, 19] and in polycrystalline samples.[20, 21] This indicates the relevance of studying the mechanisms of crystallization of boron-doped diamond.

Boron-doped diamonds can be produced at high pressures and temperatures from various starting materials. In many works, boron carbide and graphite serve as the initial components, and the synthesis is carried out at pressures above 8 GPa and temperatures 2350÷2550 °C, which exceeds the melting point of the B-C eutectic (2075 °C at normal pressure). Polycrystalline diamonds are formed as a result of liquid phase recrystallization. They have a unit cell constant increased from 3.567 Å to 3.575 Å. According to some authors,[11, 22-24] the boron content in the lattice of such diamonds can reach 4 at.%.

Several studies have demonstrated the possibility of simultaneously reducing the temperature and pressure of diamond synthesis using hydrocarbons and boron-containing starting components (it is known that the direct transition of graphite to diamond requires pressures above 12 GPa and temperatures above 2000 °C).[25] Nanodiamonds were obtained at pressures of 8-9 GPa and temperatures of about 1400 °C during the destruction of compound 9-borabicyclo [3,3,1] nonane dimer $C_{16}B_2H_{30}$.[26] Highly boron-doped micron-sized diamonds were obtained at 8 GPa and 1700 °C from mixtures of naphthalene ($C_8H_{10}$) with ortho-carborane ($C_2B_{10}H_{12}$),[27] as well as from globular nanocarbon and 1,7-di(oxymethyl)-M-carborane ($C_4B_{10}H_{16}O_2$).[28] Micron diamonds with a high (about 3 at.%) level of boron-doping were synthesized from a mixture of turbostratic globular carbon and amorphous boron.[29]

In this work the boron-doped graphite and diamond were synthesized at high pressures and temperatures from a mixture of adamantane and ortho-carborane (hereinafter referred to as carborane). It should be noted that the use of adamantane and ortho-carborane greatly reduces the influence of oxygen and other impurities on the synthesis of graphite and diamond. The obtained samples are studied with the X-ray analysis (XRD), X-ray photoelectron spectroscopy (XPS), Raman spectroscopy (Raman), and scanning electron microscopy (SEM).

**Results and Discussion**

Figure 1 shows diffraction patterns of samples obtained at a pressure of 8 GPa and a temperature of 1700 °C from: 1) carborane; 2) adamantane; 3) StartMix (adamantane and carborane mixture with B:C atomic ratio 5:95). The microphotographs of corresponding samples are given in Figure 2. Figure 1.1 and 2.1 evidence that boron carbide crystallites with a star-shaped morphology are obtained from carborane. From Figure 1.2 and 2.2 it can be seen that only graphite with lamellar particles of a round shape (up to 5 microns in diameter and up to 0.4 microns thick) is obtained from adamantane. Figure 1.3 and 2.3 show that diamond microcrystallites (up to 5 μm) and a small admixture of graphite (<5 %) are synthesized from StartMix. This result suggests that the presence of boron lowers the parameters (pressure and temperature) of diamond synthesis.

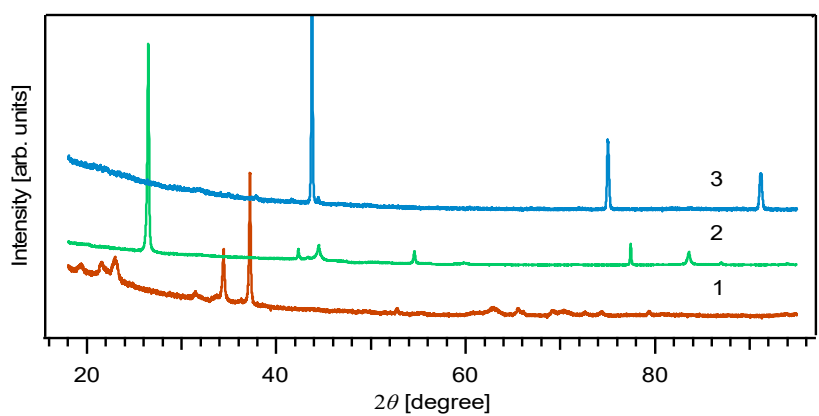

**Figure 1.** X-ray diffraction patterns of the samples obtained at 8 GPa/1700 ºC from: 1) carborane; 2) adamantane; 3) adamantane and carborane mixture (B:C atomic ratio 5:95).

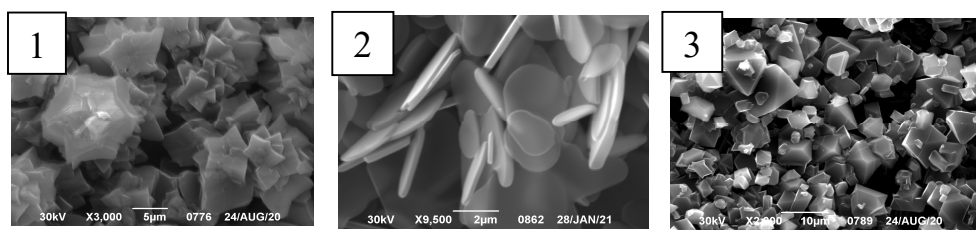

**Figure 2.** Morphology of crystals obtained at 8 GPa/1700 ºC from: 1) carborane; 2) adamantane; 3) adamantane and carborane mixture (B:C atomic ratio 5:95).

The pattern of a sample obtained from pure carborane (Figure 1.1) looks like a pattern of carbide with stoichiometry $B_{13}C_2$, in which the peaks are shifted. This substance will not be discussed in detail in here, since there are other works considering boron carbide with a carbon deficiency.[30]

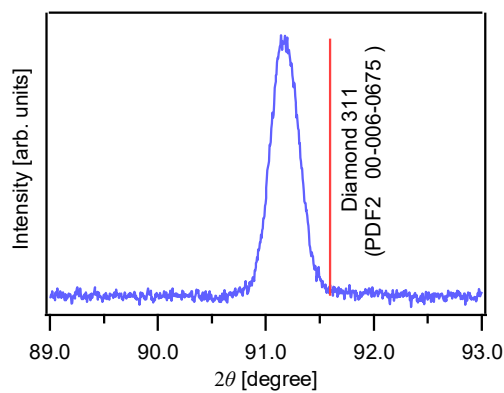

**Figure 3.** Fragment of the X-ray diffraction pattern of boron-doped diamond obtained from the adamantane and carborane mixture (B:C atomic ratio 5:95) at 8 GPa/1700 °C.

Peaks in the pattern of the boron-doped diamond obtained at 8 GPa/1700 °C from the StartMix (Figure 3) are shifted from "standard" positions. This fact is known and is associated with a change in the lattice constant upon doping with boron. Indeed, the X-ray diffraction pattern refinement gives $a = 3.5766(5)$ Å, which differs from the lattice constant of pristine diamond. The boron concentration for this case was estimated at 2÷2.5 at.%.[22]

Figure 4 shows the Raman spectrum of a boron-doped diamond obtained from the StartMix at 8 GPa/1700 °C. The spectrum is dominated by bands characteristic of a diamond with a high degree of boron-doping:[31] wide band with a maximum of about 492 cm$^{-1}$, and other bands at 1000 cm$^{-1}$, 1213 cm$^{-1}$, and 1297 cm$^{-1}$.

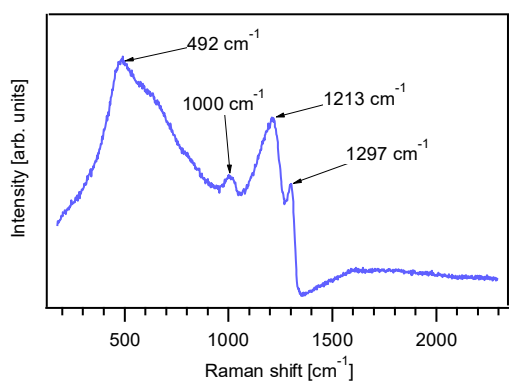

**Figure 4.** Raman spectrum of a diamond obtained at 8 GPa/1700 °C from the adamantane and carborane mixture (B:C atomic ratio 1:19).

Figure 5 shows diffraction patterns of samples obtained from StartMix at 1700 ºC and various pressures. Their analysis allows to conclude that at pressures below ~7 GPa, StartMix turns into graphite rather than diamond. In other words, the pressure-boundary of the diamond synthesis from the mixture of the adamantane and carborane lays above the 7 GPa and below 8 GPa.

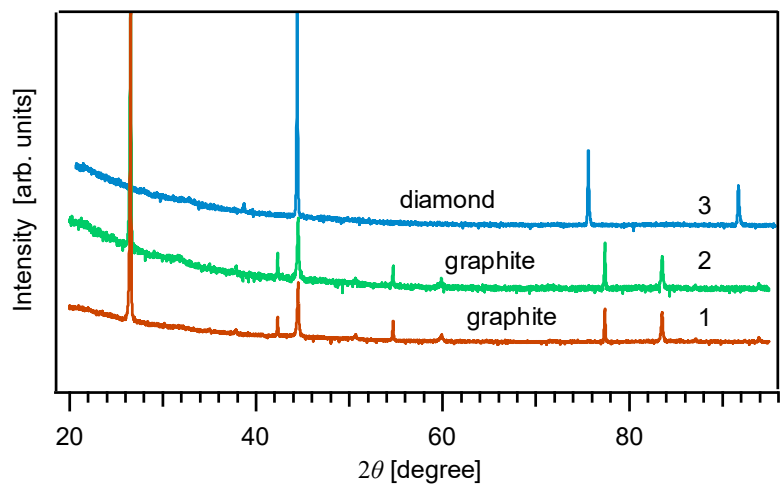

**Figure 5.** X-ray diffraction patterns of the samples obtained from the adamantane and carborane mixture (B:C atomic ratio 5:95) at 1700 ºC and: 1) 5.5 GPa; 2) 7 GPa; 3) 8 GPa.

Figure 6 shows diffraction patterns of the graphite obtained from the StartMix at 5.5 GPa and different temperatures. Figure 7 shows graphite crystals obtained at 5.5 GPa/1400 ºC from pure adamantane and from StartMix. Under the pressure of 5.5 GPa the graphitization sets in at relatively low temperatures, whereas at 1400 ºC a three-dimensionally ordered graphite is formed (Figure 6.2, 7.2). Graphite was also obtained from pure adamantane at 5.5 GPa and 1400 ºC (Figure 6.1, 7.1). The diffraction peaks of the samples obtained from the StartMix (Figure 6.2, 6.3) are shifted relative to both the "standard" positions and peaks of the sample produced from pure adamantane (Figure 6.1). This shift indicates that boron atoms replace carbon atoms in the graphite lattice sites during doping.[32] It was found that at higher temperatures (1700 ºC) the graphite obtained from the StartMix becomes more perfect. This is manifested in the narrowing of the diffraction peaks. At the same time, the X-ray peaks become even more displaced, which indicates a higher boron doping (Figure 6).

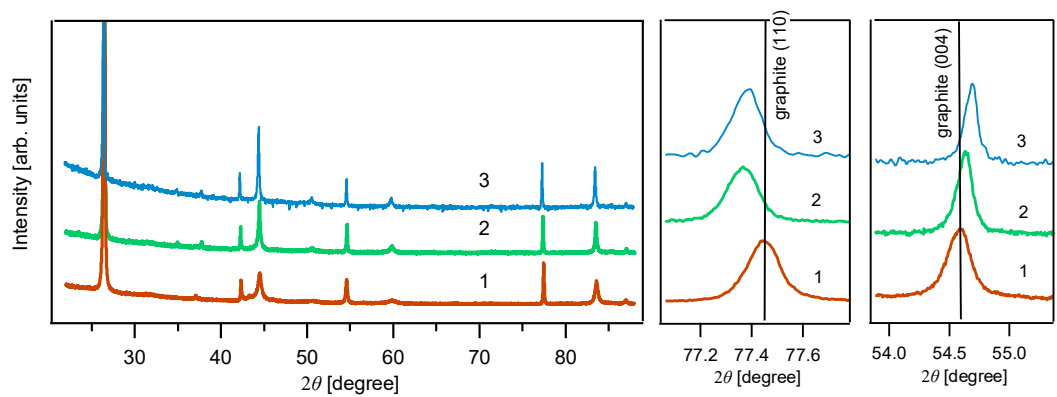

**Figure 6**. X-ray diffraction patterns of the samples obtained at 5.5 GPa from: 1) adamantane at 1400 ºC; 2) adamantane carborane mixture (B:C atomic ratio 5:95) at 1400 ºC; 3) adamantane and carborane mixture (B:C atomic ratio 5:95) at 1700 ºC.

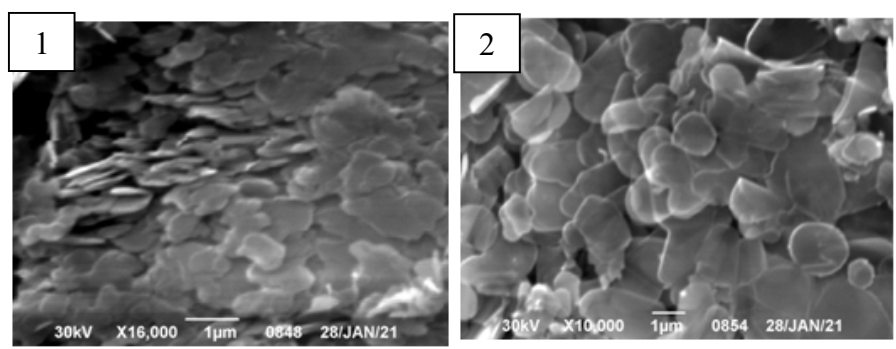

**Figure 7.** Morphology of graphite crystals obtained at 5.5 GPa/1400ºC from: 1) adamantane; 2) adamantane and carborane mixture (B:C atomic ratio 5:95).

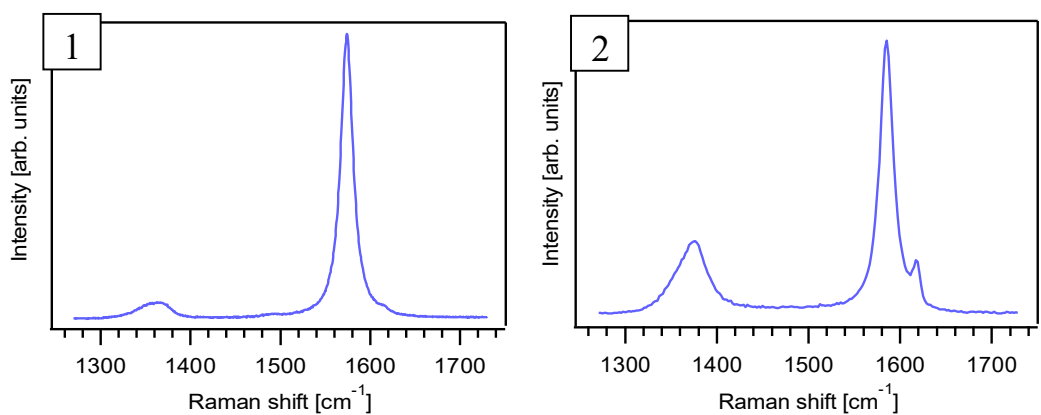

**Figure 8.** Raman spectra of the samples obtained at 5.5 GPa/1400 ºC from: 1) adamantane; 2) adamantane and carborane mixture (B:C atomic ratio 5:95).

Figure 8 shows the Raman spectra of graphite produced from adamantane and the StartMix at 5.5 GPa and 1400 ºC. Both spectra are dominated by an intense graphite line near 1590 cm$^{-1}$ (G mode), and there are also peaks centered at 1370 cm$^{-1}$ (D mode) and 1620 cm$^{-1}$ (D' mode). The increased intensity of the D and D' lines can be attributed to the growth of the structural disorder of graphite and to a high degree of boron doping.[32-34] According to our X–ray diffraction data, the incorporation of boron into graphite lattice does not violate its perfection. That is why we associate the increase in the intensity of the D and D' bands with the purely concentration effect of boron impurity on the phonon spectrum of boron-doped graphite.

The results of X-ray photoelectron spectroscopy studies of boron-doped graphite and diamond obtained from the StartMix are reproduced in Figure 9, 10 and Table 1.

The shape of the C1s spectrum for the boron-doped graphite obtained from the StartMix at 5.5 GPa/1400 ºC is typical of conventional graphite. It was approximated by a narrow asymmetric peak at 284,2 eV with FWHM (full width at half maximum) of 0,7 eV and a shake-up ($\pi - \pi^*$) satellite shifted by 6.0 eV (Figure 9.a). These parameters are consistent with C1s protocol for highly oriented pyrolytic graphite (HOPG),[35] but the binding energy (BE) value is 0.2 eV less than the values obtained for reference graphite samples without impurities of heteroatoms (quasi-single crystal graphite, HOPG). The decreased BE value can be explained by secondary shifts from carbon-boron bonds, since the carbon spectrum also contains a low-intensity peak at 282.5 eV, which is characteristic to C-B bonds.

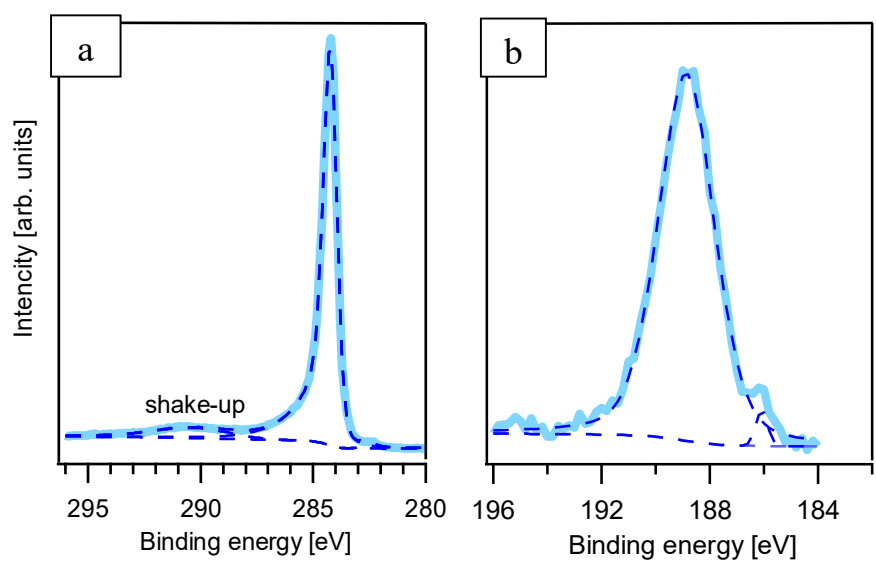

**Figure 9.** XPS spectra of boron-doped graphite obtained from the adamantane and carborane mixture (B:C atomic ratio 5:95) at 5.5 GPa/1400 ºC: a) C1s; b) B1s.

The XPS B1s spectrum (Figure 9.b) of boron-doped graphite contains a main peak at 188,7 eV (FWHM 2.5 eV) and a low-intensity peak at 186.1 eV (FWHM 0.7 eV). The peak at 188.8 eV corresponds to the B-C chemical bonds and can be attributed to boron atoms in the substitution position in graphite layers.[36] The BE of the low-intensity peak (186.1 eV) is less than BE of B1s for elemental boron (187.2 ÷ 187.7 eV). Further research is required to accurately reference it. Several configurations can be suggested for consideration: boron atoms between graphite layers, boron atoms linked to vacancies in graphite layers.[37, 38] It was suggested to associate the decrease in the BE of B1s with the boron clusters formation.[39]

**Table 1.** B1s and C1s spectral parameters:
full width at half maximum (FWHM), binding energy (BE) and fraction

| Sample | Spectral parameter | B1s | | | C1s | | | |
|---|---|---|---|---|---|---|---|---|
| | | ? | B-C (B-O) | C-B | C-C | | C-O | shake-up satellite |
| boron-doped graphite obtained from adamantane and carborane mixture at 5.5GPa/1400ºC | BE (±0.2), [eV] | 186.1 | 188.7 | 282.5 | 284/2 | | - | 290.2 |
| | FWHM (±0.05), [eV] | 0.7 | 2.5 | 0.7 | 0.7 | | - | 3.5 |
| | Fraction (±5), [%] | 3 | 97 | 1 | 90 | | - | 9 |
| boron-doped diamond from adamantane and carborane mixture at 8GPa/1700ºC | BE (±0.2), [eV] | - | 188.7 | 282.8 | 284.0 | 284.7 | 286.3 | - |
| | FWHM [±0.05], [eV] | - | 2.5 | 0.9 | 0.8 | 1.3 | 1.4 | - |
| | Fraction (±5), [%] | - | 100 | 3 | 46 | 48 | 3 | - |

The C1s spectrum of the boron-doped diamond (Figure 10.a), obtained from the StartMix at 8GP/1700 ºC was fitted by two major symmetric peaks at 284.0 eV and 284.7 eV, a peak at 286.3 eV (3 %) in the C-O bond region and a peak at 282.8 (3 %) in the carbide region (281.9 - 282.8 eV). The corresponding spectrum of B1s (Figure 10.b) is rather wide (BE - 188.7 eV, FWHM - 2.5 eV), but has no clear signs of the presence of two or more peaks.

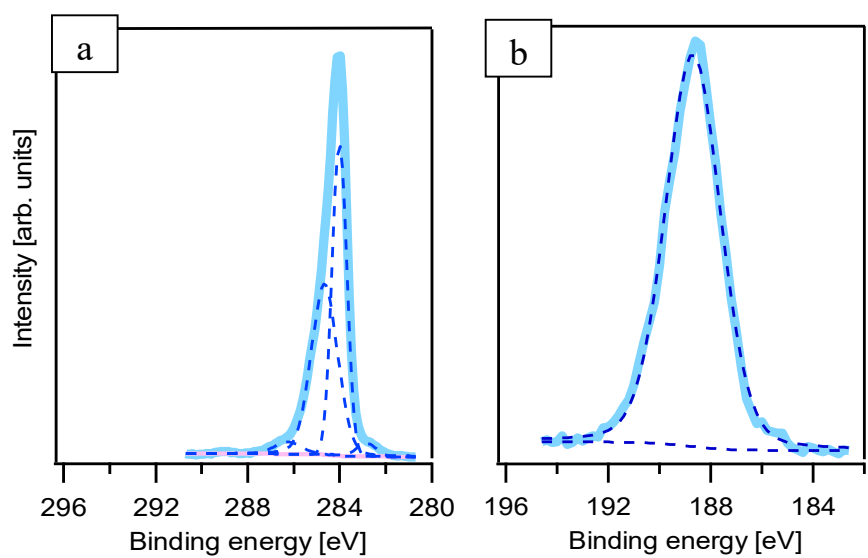

**Figure 10.** XPS spectra of boron-doped diamond obtained from the adamantane and carborane mixture (B:C atomic ratio 5:95) at 8 GPa/1700 ºC: a) C1s; b) B1s.

**Discussion**

It is known that boron and its compounds often contain a significant amount of oxygen on their surface. The use of a mixture of adamantane with carborane makes it possible to ignore its role in the synthesis of boron-doped graphite and diamond.

Experimental results evidence that adamantane-type hydrocarbons at 5.5 GPa begin to graphitize at relatively low temperatures, below 1400 °C. The addition of boron to adamantane in the form of carborane makes the resulting graphite more perfect, whose diffraction peaks are narrower (Figure 6). Calculated on the basis of X-ray data, the size of the coherent scattering regions of boron-doped graphite obtained at 5.5 GPa and 1400 °C from a mixture of adamantane-carborane, ~ 480 nm, is larger than that of pure graphite from the adamantane, ~ 270 nm. Boron-doped graphite obtained at higher temperatures (1700 °C) has even larger coherent scattering regions, ~ 720 nm.

As already mentioned, and shown in the Figure 6, an increase in temperature shifts the diffraction peaks more, which indicates an increase in boron concentration. Thus, the temperature causes a simultaneous increase in the boron content and an improvement in the resulting graphite structure.

The presence of boron and the use of high pressures reduce the graphitization temperature and increase the quality of the resulting graphite.[40, 41] Our result shows that these effects can support each other. That is, the pressure lowers the graphitization temperature of both adamantane and adamantane with carborane. But the presence of boron further improves the structure.

The replacement of carbon atoms with boron atoms leads to a decrease in the lattice constant *c* of the hexagonal unit cell and an increase in the lattice constant *a*.[32] The same effect takes place in our case, and the diffraction peaks are shifted from their "normal" positions (Figure 6). To estimate the amount of boron $X_B$, we have used the empirical relation for the dependence of the graphite lattice constant *a* on the atomic boron content ($a = 0.24612 + 0.031 X_B$) under the assumption that the B-C bond length is 0.148 nm.[42] In graphite obtained from the StartMix at 5.5 GPa/1400 °C the $X_B$ was estimated at 1.13 %, at 5.5 GPa/1700 °C $X_B$ was estimated at 1.4 % This values are rather high, taken into account that the maximal admissible boron concentration is 2.35 at.%, achieved in the boron-carbon system at 2350 °C.[43]

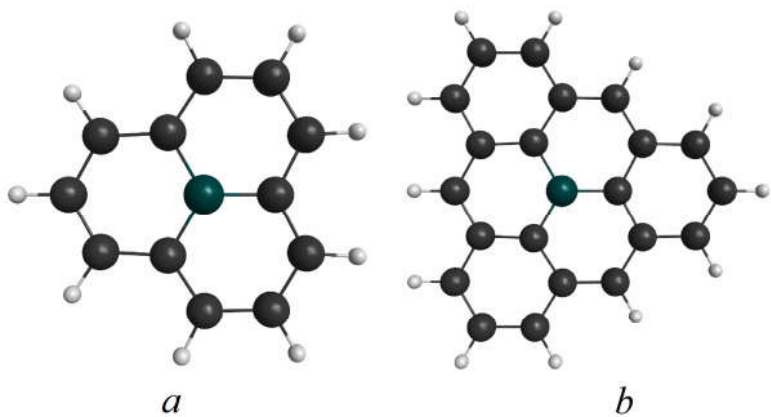

**Figure 11.** Two carbon clusters with a single substituted boron atom at the center modeling a graphene layer doped by boron: a) $BC_{12}H_9$, b) $BC_{21}H_{12}$.

To analyze the influence of doped boron atoms on the chemical bonding in a graphene layer, we have used *ab initio* quantum chemical calculations of two carbon clusters terminated by hydrogen (Figure 11). The calculations have been performed within the Hartree-Fock (HF) method implemented in GAMESS(US).[44, 45] We have used the restricted open shell version of HF for the $C_{13}H_9$ cluster ("undoped" analog of $BC_{12}H_9$, Figure 11.a) and for the $BC_{21}H_{12}$ cluster, Figure 11.b, with odd total number of electrons.

Remarkably, our calculations for both clusters show that upon substitution of the graphene layer with boron, the boron-carbon layer with the optimized geometry is not curved and remains strictly plane. This is accompanied by a change of bond lengths. We found that for the 6-31G$^{**}$ basis set, the central bond lengths increase from 1.422 Å for $C_{13}H_9$ to 1.499 Å for $BC_{12}H_9$. However, in both clusters shown in Figure 11, positions of carbon atoms within a cluster are not fully equivalent. For example, in $C_{13}H_9$ the C-C bond lengths lie in the range from 1.383 Å to 1.422 Å. Therefore, it is more instructive to compare average bond lengths (*d*), which include both the B-C and C-C bonds for the substituted case and only C-C bonds for the non-substituted one. Calculations show that *d* increases from 1.402 Å for $C_{13}H_9$ to 1.423 Å for $BC_{12}H_9$. This finding is close to the experimental observation giving the growth of *d* from 1.42 Å for the pristine graphite to 1.43 Å for the sample strongly doped with boron. The actual value of *d* for the doped sample clearly depends on the boron concentration.

Our calculations also show that the charge $Q_B$ of the boron impurity with the three neighboring carbon atoms is negative. The value of $Q_B$ is -0.326*e* according to the Mulliken population analysis and -0.572*e* according to the Löwdin population analysis (using diagonal elements of the density matrix).[43] This is in line with the general model for a pure semiconductor doped with acceptor impurities.[45] To a first approximation, the boron impurity can be represented as a carbon atom with a point negative charge *e* placed at its site, along with an additional hole delocalized in its neighborhood. In our case taking into account the cluster geometry optimization, the effective charge of the boron impurity is reduced from one to $Q_B$. This fact has important consequences: for a sufficiently large concentration of acceptor impurities, the substitutional boron atoms will experience the Coulomb repulsion of neighboring centers and therefore prefer to stay at equal distances from each other. This holds also for boron impurities lying in neighboring graphene planes where the optimal arrangement is such that the boron atoms are located under the positive charge associated with hole states and vice versa. This structure is typical for ionic crystals and results in an electrostatic energy gain and a slight decrease of interplane distances, which is confirmed by experimental observation showing that the diffraction peaks are not broadened but become noticeably shifted.

It is worth noting that this scenario leads to certain selected impurity concentrations when the arrangement of boron atoms within the graphene layer is most preferable. Such concentrations correspond to a superstructure formed by the boron atoms in graphene sheets. At small concentrations (< $c_0$) boron atoms being far away from each other do not experience the Coulomb repulsion which is very small. In that case no superstructure of boron atoms is formed. At large concentrations (> $c_0$) the impurity positions become regular. Then two periodic structures should be considered: the first is formed by the substitutional boron atoms, the second is given by the original carbon lattice. The two structures can have common translations (a), or stay irregular (b). The latter case (b) corresponds to irregular boron doping implying a glass state or quasicrystal. We next consider the case (a) which is most stable and energy preferable. For the regular structure one can further define a supercell with a single boron impurity, whose basis vectors are *n* times larger than the initial basis vectors of graphene. (Notice that the two basis vectors of the supercell should be equal for otherwise the distances between neighboring boron atoms will be different.) One can show that in that case the concentration of the boron atoms is $1/(2n^2)$ while the nearest B-B distance $d_{B-B}$ is given by $3^{1/2} \times d \times n$, where $d$ = 1.43 Å is the averaged C-C bond length between two neighboring carbon atoms in the graphene lattice doped with boron. The results for superstructures with *n* = 2, 3, … 10 are reproduced in Table 2.

**Table 2.** Estimated repulsion energy *V* (eV) between two boron impurities in the graphene layer in dependence of the supercell index *n* and the boron concentration (%). $d_{B-B}$ (Å) is the closest distance between neighboring boron atoms, $Q_B$ = -0.326$e$.

| *n* | 2 | 4 | 5 | 6 | 7 | 8 | 9 | 10 |
|---|---|---|---|---|---|---|---|---|
| Concentration [%] | 12.5 | 3.13 | 2.0 | 1.39 | 1.02 | 0.781 | 0.617 | 0.50 |
| $d_{B-B}$ [Å] | 4.95 | 9.91 | 12.38 | 14.86 | 17.34 | 19.81 | 22.29 | 24.77 |
| *V* [eV] | 0.309 | 0.154 | 0.124 | 0.103 | 0.088 | 0.077 | 0.069 | 0.062 |

Inspection of Table 2 shows that the repulsion changing considerably at small *n*, displays only small changes at relatively large **n** = 9, 10, … . We can roughly estimate the critical concentration $c_0$ by comparing the repulsion with a characteristic energy of lattice displacements described by the Debye temperature $\Theta_D$. Taking $\Theta_D \approx$ 1000 K,[46] we obtain that the repulsion becomes smaller the elastic energy ($\Theta_D$) at **n** ≥ 8. Therefore, the critical concentration corresponds to the value of $c_0$ = 0.8 %.

From the XPS data, it follows that the main C1s peak, apparently associated with C-C bonds, is located at ~284.2 eV. This is slightly lower than the usual value recorded for graphite, 284.4 eV, when the device is calibrated using the spectra of gold Au4f (83.96 eV) and copper Cu2p3 (932.62 eV). A similar shift of the C1s peak was observed in boron-doped nanotubes.[47] This result correlates with our observations, reflecting the fact that the energy and length of the C-C bonds change with boron substitution.

We believe that the real mechanism of boron promotion of diamond synthesis is related to the structure of boron-doped graphite, which mediates the transition of adamantane to diamond. That is, there is an intermediate graphite (or graphite-like) phase doped with boron. The actual mechanism could involve destabilizing the graphite lattice and/or lowering the energy barrier.

Below we briefly discuss the issue and argue that it is mainly boron that facilitates the synthesis of diamond and its influence can be separated from the effects caused by surface hydrogen and oxygen.

It is worth noting that the transition of pure graphite to diamond requires pressures of the order of 12÷25 GPa and temperatures above 2000 ºC.[25] Although at present there is no consent on the role of hydrogen, Davydov *et al* have reported that the synthesis of diamonds from polycyclic aromatic hydrocarbons (PAHs: naphthalene, anthracene, pentacene, perylene and coronene) takes place at 8 GPa and elevated temperatures.[48] However, in other works this has not been observed.[see references in 48] Nevertheless, a mechanism has been proposed which can facilitate the transformation of the graphite lattice into diamond through its surface hydrogenation.[49]

As noted here, pure adamantane (which contains hydrogen) at 8 GPa/1700 ºC was converted to graphite, while a mixture of adamantane and carborane (which additionally contained boron) was converted to boron-doped diamond. Thus, hydrogen alone does not promote diamond synthesis. The fact that hydrogen does not play a key catalytic role was proved by the synthesis of diamond from a mixture of fullerene and amorphous boron at 8 GPa and the temperatures ~1400 ºC.[50] It was reported that even in the absence of hydrogen, boron-doped graphite transforms into diamond at 7.5÷8.0 GPa and 1600÷1650 ºC.[29] That is, there is a clear catalytic effect cased solely by boron. Present work confirms that the promotional effect of boron on the synthesis of diamond can be considered separately from the effect of oxygen, since the initial substances did not contain it.

**Conclusions**

The experimental results obtained show that high-pressure methods are effective for producing graphite and diamond with a high degree of boron doping from hydrocarbons and carboranes. In particular, perfect boron-doped graphite with a coherent scattering region size $L_a$ ~ 500 nm can be obtained from an adamantane-carborane mixture (with the atomic B to C ratio 5:95) at 5.5 GPa and 1400 ºC. It is shown that 1.13 at.% of boron atoms substitute carbon in the graphite lattice. This amount of boron is quite high, since the maximal solubility of boron in graphite is 2.35 at.% reached at 2350 ºC.[43]

It is known that the presence of boron in the starting material and the use of high pressures reduce the graphitization temperature. Our result shows that these effects can support each other.

The X-ray diffraction patterns of boron-doped graphite allow us to suggest that its lattice is quite perfect even at high boron concentrations, with a regular and possibly periodic arrangement of boron atoms in graphite layers. Our *ab initio* calculations of small planar carbon clusters show that they remain flat upon replacing carbon by boron, with the average length of the B-C and C-C bonds increased by 0.02 Å, which is consistent with the X-ray diffraction data. It also follows from the calculations that an effective negative charge formed around each boron site, leads to the Coulomb repulsion between neighboring impurities and, thus, can explain their regular equidistant position in the graphite layer.

Carbon C1s XPS spectrum of the boron-doped graphite obtained at 5.5 GPa/1400 ºC from the adamantane and carborane mixture (with the atomic B to C ratio 5:95) has a profile characteristic of pristine graphite, but the main peak at 284.2 eV is shifted by minus 0.2 eV relative to its "standard" position. This shift can be explained by the secondary effect of carbon-boron bonds. The peak at 188.7 eV in the B1s spectrum corresponds to the C-B bonds of boron atoms in the substitutional sites of the graphite layers.

It was proved that boron has a catalytic effect on the synthesis of diamond. In particular the adamantane-carborane mixture (with the atomic B to C ratio 5:95) transformed into boron-doped diamond at 8 GPa and 1700 ºC (the direct conversion of pure graphite into diamond requires 12÷25 GPa and 2300÷2500 °C).[23] Our results and literature data indicate that the surface hydrogen and oxygen do not possess independent key catalytic properties. This, apparently, does not exclude the possibility that they can influence some other factors, such as the growth rate of diamond crystals and their faceting.

**Experimental**

The starting components, adamantane $C_{10}H_{16}$ (Sigma-Aldrich, 99.5 %) and ortho-carborane $B_{10}H_{10}C_2H_2$ (Yuanli chemical group Co. LTD, 98.0 %), were mixed in an agate mortar in hexane with the imposition of ultrasound so that the atomic B to C ratio was 5:95 (referred to as StartMix). This StartMix was dried in air at room temperature.

Thermobaric treatment was carried out in "toroid" apparatus.[49] For this, a tantalum ampoule with a sample was placed inside a graphite heater and then into a container made of lithographic stone. When pressed with profiled carbide anvils, the container experienced plastic flow and transferred pressure to the working area. The pressure was calibrated using the generally accepted technique of recording phase transitions of reference substances.[52] The temperature was measured with thermocouples.

X-ray diffraction spectra were obtained using a Huber Imaging Plate Guinier camera G670 (Cu $K_{\alpha1}$).

The photoelectron spectra were recorded with PHI 5500 VersaProbeII spectrometer. A monochromatic Al Kα source (hν = 1486.6 eV) operated at 50 W and 200 μm beam diameter. The binding energy (BE) scale was calibrated against reference metals: Au4f – 83.96 eV, Ag3d – 368.27 eV and Cu2p3/2 – 93.62 eV. The binding energy of carbon C1s was 284.4 ± 0.05 eV for graphite samples. PHI relative sensitivity factors corrected by Transmission Function were used in the atomic concentration calculation.

Raman spectra were recorded with a confocal Raman spectrometer LabRam HR800 (excitation laser wavelength 473 nm).

Scanning electron microscopy was performed on a JSM-6390 JEOL setup.

**Conflict of Interest**
The authors declare no conflict of interest.

**Acknowledgements**
XPS investigations were performed using equipment of the Joint Research Center «Material Science and Metallurgy» NUST MISIS.